\documentclass[12pt,aps,showpacs,eqsecnum,nofootinbib,floatfix]{revtex4}
\usepackage{latexsym}
\usepackage{graphicx}
\usepackage{subfigure}
\usepackage{bbding}
\usepackage{pifont}
\usepackage{multirow}
\usepackage{epsfig}
\usepackage{CJK}
\usepackage{amsmath}
\usepackage{amsfonts}
\usepackage{amssymb}
\usepackage{color}
\usepackage{mathrsfs}

\def\be{\begin{equation}}
\def\ee{\end{equation}}
\def\ba{\begin{eqnarray}}
\def\ea{\end{eqnarray}}
\def\bs{\begin{split}}
\def\es{\end{split}}
\def\no{\nonumber}

\newcommand{\omits}[1]{}

\begin{document}
\begin{CJK*}{GBK}{song}

\title{The phase transition of higher dimensional Charged black holes}

\author{Xiongying Guo, Huaifan Li~\footnote{Email: huaifan.li@stu.xjtu.edu.cn}, Lichun Zhang, Ren Zhao}

\medskip

\affiliation{ Institute of Theoretical Physics, Shanxi Datong University,
Datong 037009, China}
\affiliation{ Department of Physics, Shanxi Datong University,
Datong 037009, China}

\begin{abstract}

In this paper,we have studied phase transitions of higher dimensional charge
black hole with spherical symmetry.  we calculated the local energy and local
temperature, and find that these state parameters satisfy the first law of thermodynamics
. We analyze the critical behavior of black hole thermodynamic system
by taking state parameters $(Q,\Phi)$ of black hole thermodynamic
system, in accordance with considering to the state parameters $(P,V)$ of Van der Waals system respectively.
we obtain the critical point of black hole thermodynamic system, and find the critical point is
independent of the dual independent variables we selected. This result for asymptotically flat
space is consistent with that for AdS spacetime, and is intrinsic property of
black hole thermodynamic system.

\textbf{Keywords}:de Sitter spacetime, the effective
thermodynamic quantities, phase transition, critical phenomena

\end{abstract}

\pacs{04.70.-s, 05.70.Ce}

\maketitle

\bigskip

\section{Introduction}

In recent years, the study of critical behavior of black holes has
still received a lot of attention. Particularly, the idea of including the variation of the cosmological constant $\Lambda$ in
the first law of black hole thermodynamics has attained increasing attention.
Matching the thermodynamic quantities with the ones in usual
thermodynamic system, the critical behavior of black holes can be
investigated and the phase diagram like the van der Waals
vapor-liquid system, critical exponents and Clapeyron equations can be obtained
~\cite{RBM,Dolan3,RBM2,Cvetic,Cai1,ma1,Hendi1,ma2,wei1,wei3,
zou3,Hendi2,Hendi4,Zhao1}. For the de Sitter spacetime, the thermodynamics properties
and phase transition of black hole with the method of the equivalence
quantities has been investigated, and it is shown that the result is similar
with AdS black hole~\cite{Urano,zhao2,zhao3,zhao4,zhao5,zhao6}.
This helps to further understand black hole entropy, temperature, heat capacity, et.al,
and it is also very important to improve the self-consistent
geometric theory of black hole thermodynamics.

As is well known, there are Hawking radiations for black holes in asymptotically flat
spacetime and non-asymptotically flat ones. Because the heat
capacity of black hole in asymptotically flat spacetime is negative, this black hole in
asymptotically flat space is thermodynamically unstable. We verify the thermodynamic stability of black holes
firstly should verified the black hole
is thermodynamics stable to correctly study their thermodynamic properties and phase transitions.

In order to restore thermodynamic stability so that equilibrium
thermodynamics and the phase structure can be studied, we must consider the whole systems
that include not only the black hole under consideration but also their environment~\cite{York,Braden}.
As self-gravitating systems are spatially inhomogeneous, which is different from the
usual thermodynamic system. Any specification of such system requires not just thermodynamic
quantities of interest but also the place at which they take the specified values. In this paper,
we have studied the phase transition of higher dimensional charged black hole with spherical symmetry.
To solve this problem we can place the black hole inside a finite concentric spherical
cavity, whose radii is fixed and larger than the one of black hole. The temperature is
fixed on the surface of the cavity, which could be physically realized by placing
a heat bath around the cavity. We will keep the charge inside the cavity also
fixed. This will define a canonical ensemble~\cite{Andrew,Carlip,Capela,lu1,lu2,Eune}.
We will study the phase structure and the thermodynamic properties
of the various dimensional charged black hole in this ensemble.

Refs.~\cite{JXM2,JXM3,JXM4,JXM5,JXM6,zhao7,Jing,Ali1,Day,Kastor} have investigated the critical
behavior of various black hole in AdS spacetime with Ehrenfest scheme. They found that
the phase transition of black hole is the continuous one in AdS spacetime, and the thermodynamic
quantities of the critical point satisfies the Ehrenfest equation. Refs.~\cite{Suresh,Ali,Tharanath,Ali2,Mahapatra,Caceres}
have obtained the same conclusion by studying the thermodynamics and state
space geometry of black hole in AdS space. Moreover, it is interesting to study
the thermodynamics behavior of the ensemble, which we construct for asymptotical flat space,
and identify whether the ensemble meet with the Ehrenfest equation. We obtain the
phase diagram for the thermodynamics quantities$(Q,\Phi)$ at the critical point, and the result similar to that
in AdS black hole. The thermodynamic quantities satisfy the Ehrenfest equation, so the corresponding phase transition
is a continuous phase transition.

The paper is arranged as follows: In Sec.2 we first review the higher
dimensional charged black hole in asymptotically flat space, and give
the quasilocal thermodynamics quantities of the canonical ensemble. In Sec.3
the critical behavior in different dimensional charged black hole is
investigated. Finally, the paper ends with a brief conclusion. (we use the
units $G_d=\hbar =k_B =c=1)$

\section{The thermodynamical quantity of the charged black hole with the ensemble theory in higher dimension}

The solution for charged black hole in spacetime dimensions with $d>3$ reads
\be
\label{eq1}
ds^2 = - Vdt^2 + \frac{dr^2}{V} + r^2d\Omega _{d - 2}^2 ,
\ee
where the $V(r)$ functions entering the metric are given by
\be
\label{eq2}
\begin{split}
V(r) = 1 - \frac{m}{r^{d - 3}} + \frac{q^2}{r^{2(d - 3)}}.
\end{split}
\ee
Here the parameters $m$ is related to the ADM mass($M$) of black holes,
\ba
\label{eq3}
\begin{split}
M = \frac{(d-2)\omega _{d-2}}{16\pi}m;
\quad
\omega _{d - 2} = \frac{2\pi ^{\textstyle{{d - 1} \over 2}}}{\Gamma
(\textstyle{{d - 1} \over 2})}
\end{split}
\ea
where $\omega _{d - 2}$ is the volume of unit $(d-2)$ sphere. The
parameter $q$ is related to the electric charged $Q$ as
\begin{equation}
\label{eq4}
Q = \frac{\sqrt {2(d - 2)(d - 3)} }{8\pi }\omega _{d - 2} q.
\end{equation}
The entropy of the system is given by
\begin{equation}
\label{eq5}
S = \frac{\omega _{d - 2} }{4}r_ + ^{d - 2}
\end{equation}
where $r_ + $ is the radius of the outer event horizon defined by the
condition $V(r_ + ) = 0$. Let us first consider the cavity as a boundary with
a radius $r_B$ to study quasilocal thermodynamics along the line of the
procedure in Ref.~\cite{Braden}. Then, the local temperature measured at the boundary
is given by~\cite{Braden,Andrew,Carlip,Capela,lu1,lu2,Eune}
\begin{equation}
\label{eq6}
T_{loc} = \frac{T}{\sqrt {V(r_B )} } = \frac{(d - 3)(r_ + ^{2(d - 3)} -
q^2)}{4\pi \sqrt {V(r_B )} r_ + ^{2d - 5} }.
\end{equation}
For fixed charge $Q$, the entropy calculated from the first law of
thermodynamics is
\begin{equation}
\label{eq7}
S = \int {\frac{dM}{T}} .
\end{equation}
Applying the first law of thermodynamics, the total thermodynamic internal
energy within the boundary $r_B $ is obtained as
\begin{equation}
\label{eq8}
E_{loc} = \int\limits_{M_0 }^M {T_{loc} } dS = \int\limits_{M_0 }^M
{\frac{T_{loc} }{T}} dM = \frac{(d - 2)\omega _{d - 2} }{8\pi }r_B^{d - 3}
\left( {\sqrt {V(B_0 )} - \sqrt {V(r_B )} } \right).
\end{equation}
Take the $V(B_0 ) = 1$ at the boundary of cavity~\cite{lu1}, we can obtain
\be
\label{eq9}
E_{loc} = \frac{(d - 2)\omega _{d - 2} }{8\pi }r_B^{d - 3} \left( {1 - \sqrt
{V(r_B )} } \right).
\ee
where
\be
\label{eq10}
V(r_B ) = \left( {1 - \frac{r_ + ^{d - 3} }{r_B^{d - 3} }} \right)\left( {1
- \frac{q^2}{r_ + ^{d - 3} r_B^{d - 3} }} \right).
\ee
So we can define the reduction quantities for corresponding,
\ba
x &=& \frac{r_ + ^{d - 3} }{r_B^{d - 3} },
\quad
\tilde {q} = \frac{q}{r_B^{d - 3} },
\quad
b_q (x) = \frac{(d - 2)\beta _B }{8\pi r_B }, \no  \\
h &=& \frac{(d - 2)\omega _{d - 2} }{8\pi }r_B^{d - 3} \left( {1 - \sqrt
{\left( {1 - x} \right)\left( {1 - \frac{\tilde {q}^2}{x}} \right)} }
\right), \\
S &=& \frac{\omega _{d - 2} }{4}r_B^{d - 2} x^{\textstyle{{d - 2} \over {d -
3}}}. \no
\ea
For $r_ + > q$, $r_B > r_ + $, we get $q < x <1$. The states functions in the
cavity satisfy the the first law of thermodynamics system~\cite{Braden,zhao8},
\be
\label{eq11}
dE_{loc} = T_{loc} dS + \Phi dQ + \sigma dA.
\ee
where
\be
\label{eq12}
S = \frac{\omega _{d - 2} }{4}r_ + ^{d - 2} ,
\quad
A = \omega _{d - 2} r_B^{d - 2} ,
\ee
From the Eqs.(\ref{eq9}) and (\ref{eq11}), we can obtain
\ba
\label{eq13}
\left( {\frac{\partial E_{loc} }{\partial S}} \right)_{Q,A}&=& T_{loc}=
\frac{(d - 3)\left( {1 - \frac{q^2}{r_ + ^{2(d - 3)} }} \right)}{4\pi r_ +
\left( {1 - \frac{r_ + ^{d - 3} }{r_B^{d - 3} }} \right)^{1 / 2}\left( {1 -
\frac{q^2}{r_ + ^{d - 3} r_B^{d - 3} }} \right)^{1 / 2}}, \no \\
\left( {\frac{\partial E_{loc} }{\partial Q}} \right)_{S,A}&=& \Phi=
\frac{(d - 2)q\left( {1 - \frac{r_ + ^{d - 3} }{r_B^{d -
3} }} \right)}{\sqrt {2(d - 2)(d - 3)} r_ + ^{d - 3} \left( {1 - \frac{r_ +
^{d - 3} }{r_B^{d - 3} }} \right)^{1 / 2}\left( {1 - \frac{q^2}{r_ + ^{d -
3} r_B^{d - 3} }} \right)^{1 / 2}} , \\
\left( {\frac{\partial E_{loc} }{\partial A}} \right)_{r_ + ,Q}&=&\sigma =
\frac{(d - 3)}{8\pi r_B }\left[ {1 - \frac{1 - \frac{r_ + ^{d - 3} }{2r_B^{d
- 3} } - \frac{q^2}{2r_ + ^{d - 3} r_B^{d - 3} }}{\left( {1 - \frac{r_ + ^{d
- 3} }{r_B^{d - 3} }} \right)^{1 / 2}\left( {1 - \frac{q^2}{r_ + ^{d - 3}
r_B^{d - 3} }} \right)^{1 / 2}}} \right]. \no
\ea

\section{the critical effective in charged black hole in cavity}
For the thermodynamic ensemble, we take the sate parameters $T_{loc},S$ as the independent variables to study
the critical behaviors of  thermodynamics ensemble. When these state parameters $Q$ and $r_B$ are invariable quantities,
the critical value of black hole radii $r_B$, black hole event horizon $r_+$ and the black hole temperature are decided
by,
\be
\label{eq14}
\left( {\frac{\partial T_{loc} }{\partial r_ + }} \right)_{r_B,Q} = 0,
\quad
\left( {\frac{\partial ^2T_{loc} }{\partial r_ + ^2 }} \right)_{r_B,Q} =
0.
\ee
We can calculate the position of the critical points in different dimensional spacetime. The results are shown in Table~1(we take $r_B=1$),
\begin{table}[!htbp]
\begin{center}
\begin{tabular}{|c|c|c|c|c|}
\hline
$ d $& $r_+^c$& $Q^c$& $T_{loc}^c $& $\Phi^c $\\
\hline
4&0.527864&0.236068&0.185589&0.324920\\
\hline
5&0.610537&0.320704&0.301913&0.221054\\
\hline
6&0.661764&0.378834&0.404099&0.177014\\
\hline
7&0.698081&0.404347&0.500148&0.151505\\
\hline
\end{tabular}
\label{tab1}
\end{center}
\caption{Numerical solutions for $r_+^c$, $Q^c$, $T_{loc}^c $ and $\Phi^c $ for given values of $d=4,~5,~6,~7$ respectively.}
\end{table}

\begin{figure}[!htbp]
\center{\subfigure[~$ d=4 $] {
\includegraphics[angle=0,width=7cm,keepaspectratio]{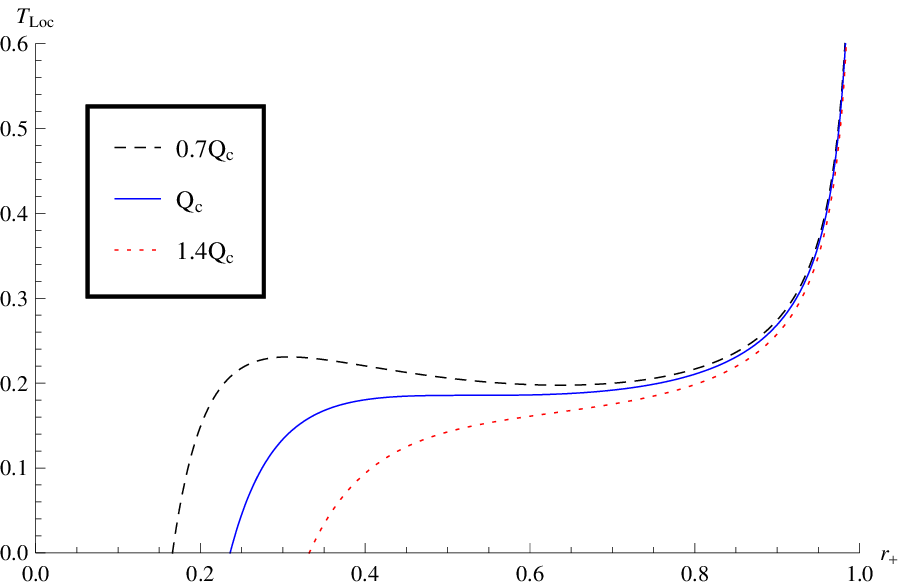}}
\subfigure[~$ d=7 $] {
\includegraphics[angle=0,width=7cm,keepaspectratio]{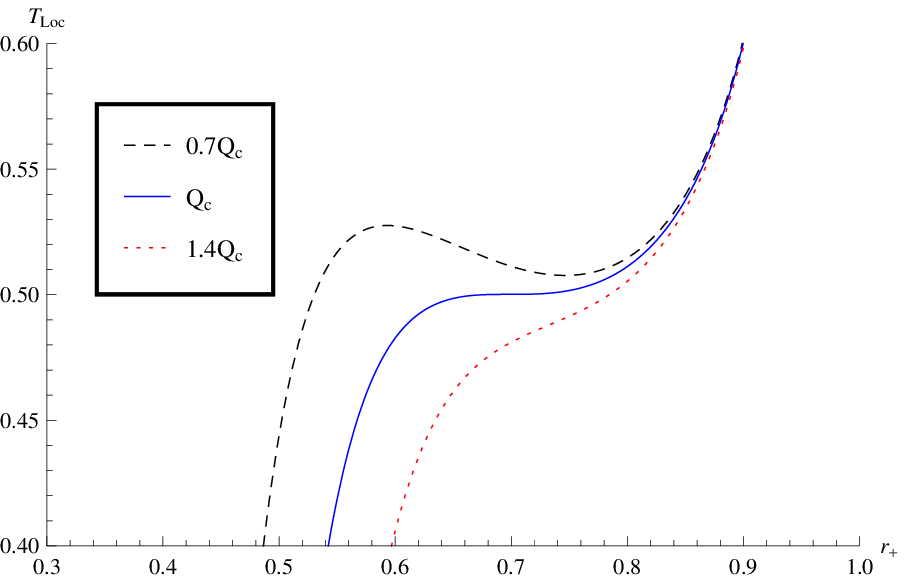}}
\caption[]{\it The $T_{loc}-r_+$ curves for $d=4,7$ respectively. From top to the bottom the curves correspond to the charge
$0.7Q_c$, $Q_c$, $1.4Q_c$.}}
\label{pr}
\end{figure}
Table.1 shows the critical values of $r_+^c$, $Q^c$, $T_{loc}^c $ increase as the spacetime dimensional $d$ increase, and
the $\Phi^c$ decreases as the spacetime dimensional $d$ increases. Fig.1 shows $T_{loc}$ curves with black hole radii for different $d$.
we can find that there is a phase transition near critical point with the charge $Q \leq Q_c$. The critical temperature $T_{loc}$ increases as the spacetime dimension $d$ increases.
From the heat capacity of system
\be
\label{eq15}
C_Q = T_{loc} \left( {\frac{\partial S}{\partial T_{loc} }} \right)
\ee
we can find that the instability state will appear with the charge of spacetime is littler the critical charge.
By using Maxwell's equal area law we discover the possible two-phase coexistence curves in the process of phase
transition~\cite{zhao8,JXM7}.

Next, let us calculate the free energy of the black hole in order to study phase transition between the black holes
and the hot flat space~\cite{Carlip,Eune}.
\be
\label{eq16}
F = E_{loc} - T_{loc} S.
\ee
We can plot the relations curve for the free energy $F$ and the local temperature $T_{loc}$, with $r_B=1$, and $Q$ takes values near critical charge.
\begin{figure}[!htbp]
\center{\subfigure[~$ d=4 $] {
\includegraphics[angle=0,width=7cm,keepaspectratio]{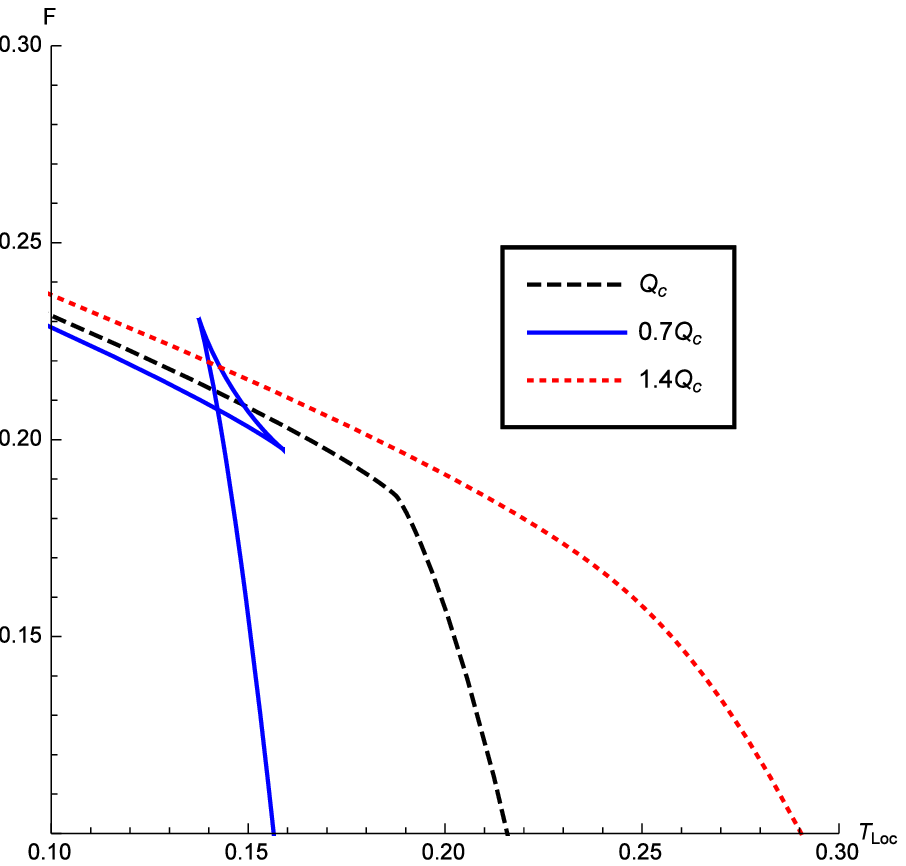}}
\center{\subfigure[~$ d=6 $] {
\includegraphics[angle=0,width=7cm,keepaspectratio]{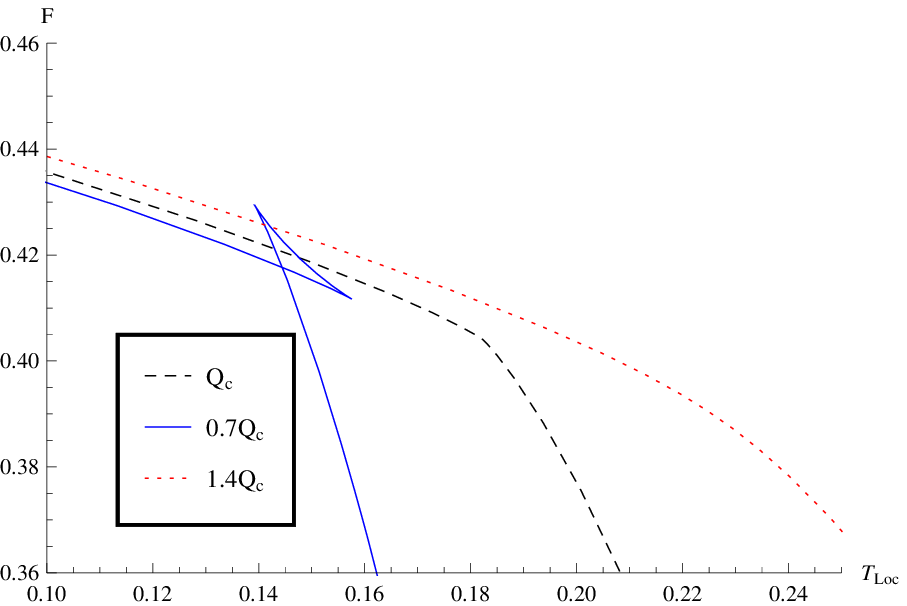}}
\subfigure[~$ d=7 $] {
\includegraphics[angle=0,width=7cm,keepaspectratio]{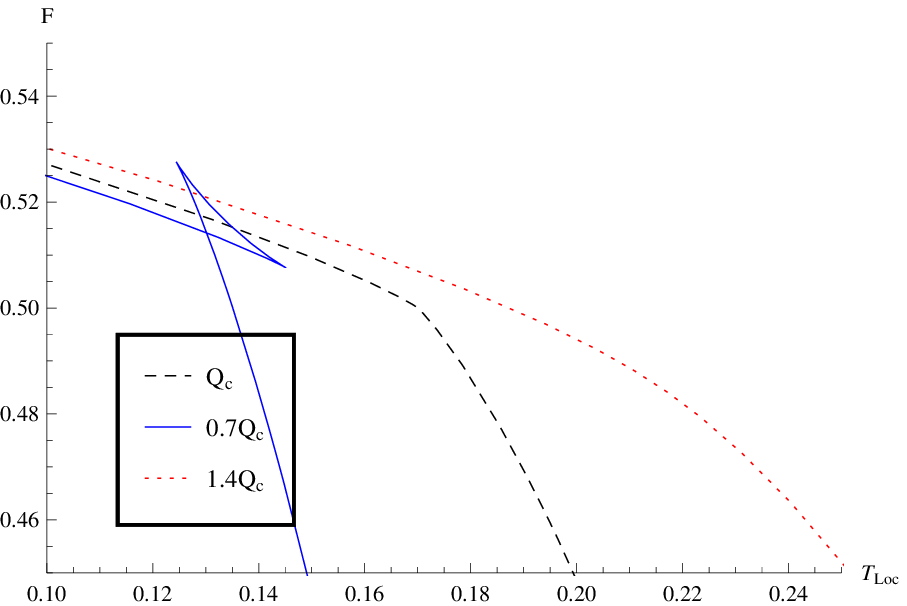}}
\caption[]{\it The $F-T_{loc}$ curves for $d=4, 6, 7$ respectively. From top to the bottom the curves correspond to the effective temperature
$0.7Q^c$, $Q^c$, $1.4Q^c$.}}}
\label{FT1}
\end{figure}
Fig.2 shows the system is the two-phase coexistence state, when the charge of black hole is smaller than the critical
charge $Q_c$. This result is consistent with Fig.1. From Table 1, it is found that the critical temperature and
critical charge increase as the spacetime dimension increases. This result is consistent with the conclusion in Refs.~\cite{RBM2} for higher dimensional AdS black hole.

Further, We can plot the curve for the free energy $F$ and the local temperature $Q$, with $r_B=1$, and $T_{loc}$ take
values near critical temperature in Fig.3. From the Fig.3, we can see that there is a phase transition at the quasilocal
temperature smaller than the critical quasilocal temperature in different dimensionas. The results is consistent with the profile
of $F-T_{loc}$ in Fig.2.
\begin{figure}[!htbp]
\center{\subfigure[~$ d=4 $] {
\includegraphics[angle=0,width=5cm,keepaspectratio]{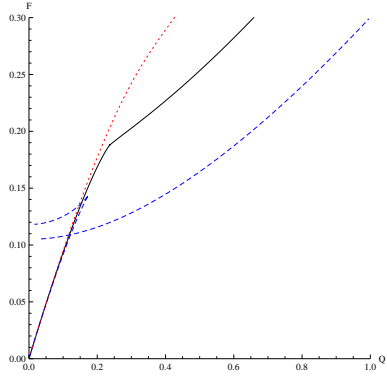}}
\center{\subfigure[~$ d=6 $] {
\includegraphics[angle=0,width=5cm,keepaspectratio]{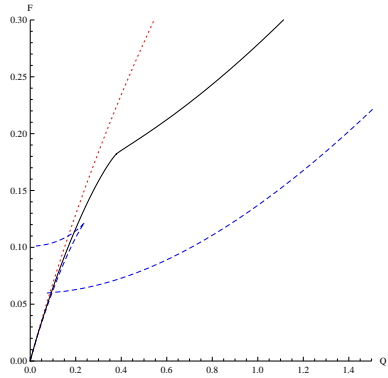}}
\subfigure[~$ d=7 $] {
\includegraphics[angle=0,width=5cm,keepaspectratio]{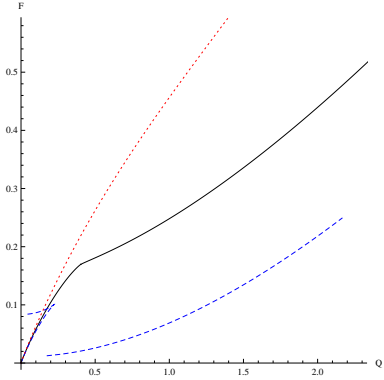}}
\caption[]{\it The $F-Q$ curves for $d=4, 6, 7$ respectively. From top to the bottom the curves correspond to the effective temperature
$0.7T_{loc}$, $T_{loc}$, $1.4T_{loc}$.}}}
\label{FT2}
\end{figure}

From the (\ref{eq11}), we know that we can select the independent variables $(Q,\Phi)$ or $(\sigma, A)$
for the black hole thermodynamic ensemble. When taking the $r_B$ and $T_{loc}$ as constant, we select the dual independent
variables $(\Phi, Q)$, the critical position is given by the following conditions,
\ba
\label{eq17}
\left( {\frac{\partial Q}{\partial \Phi }} \right)_{T_{loc} ,r_B } &=&
\frac{\left( {\frac{\partial T_{loc} }{\partial r_ + }} \right)_Q }{\left(
{\frac{\partial \Phi }{\partial Q}} \right)_{r_ + } \left( {\frac{\partial
T_{loc} }{\partial r_ + }} \right)_Q - \left( {\frac{\partial T_{loc}
}{\partial Q}} \right)_{r_ + } \left( {\frac{\partial \Phi }{\partial r_ +
}} \right)_Q } = \frac{\left( {\frac{\partial T_{loc} }{\partial r_ + }}
\right)_Q }{\frac{\partial (\Phi ,T_{loc} )}{\partial (Q,r_ + )}} = f(r_ +
,Q) = 0, \no \\
\left( {\frac{\partial ^2\Phi }{\partial Q^2}} \right)_{T_{loc} ,r_B } &=&
\left( {\frac{\partial f(r_ + ,Q)}{\partial Q}} \right)_{T_{loc} ,r_B } =
0.
\ea
We can plot the curve for the free energy $\Phi$ and the local temperature $Q$ in Fig.4,
with $r_B=1$, and $T_{loc}$ take values near critical temperature.
\begin{figure}[!htbp]
\center{\subfigure[~$ d=4 $] {
\includegraphics[angle=0,width=5cm,keepaspectratio]{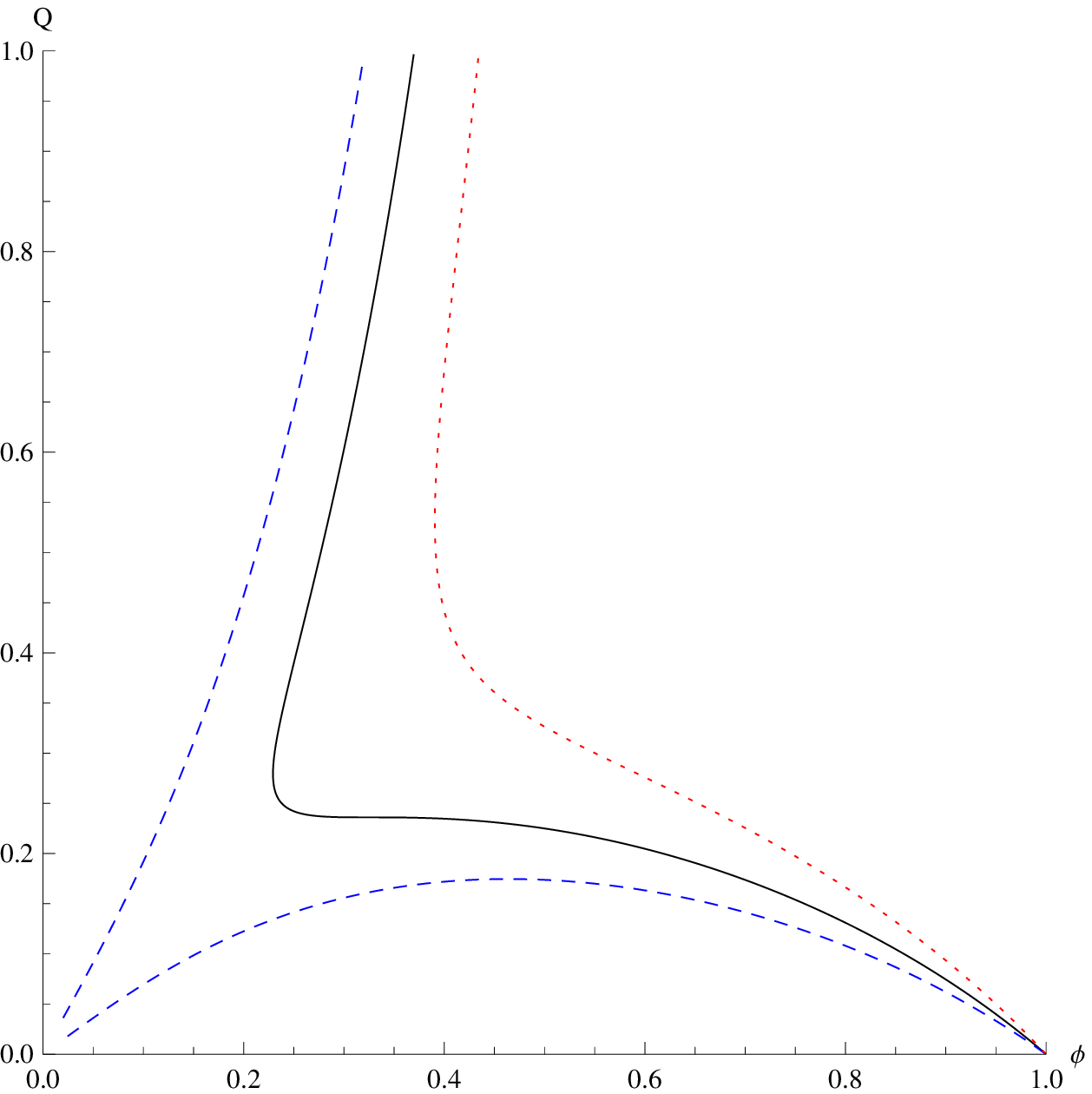}}
\center{\subfigure[~$ d=6 $] {
\includegraphics[angle=0,width=5cm,keepaspectratio]{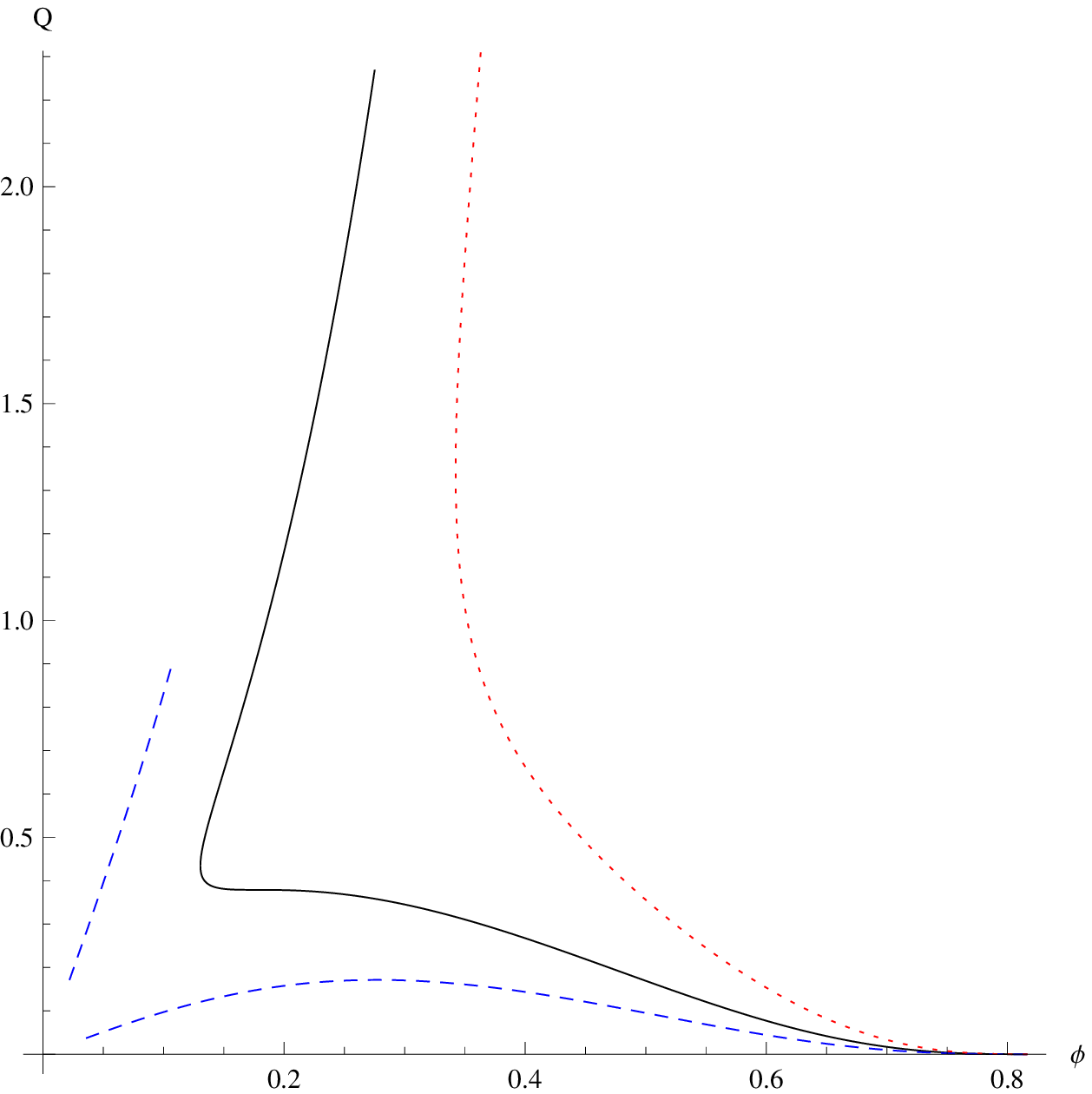}}
\subfigure[~$ d=7 $] {
\includegraphics[angle=0,width=5cm,keepaspectratio]{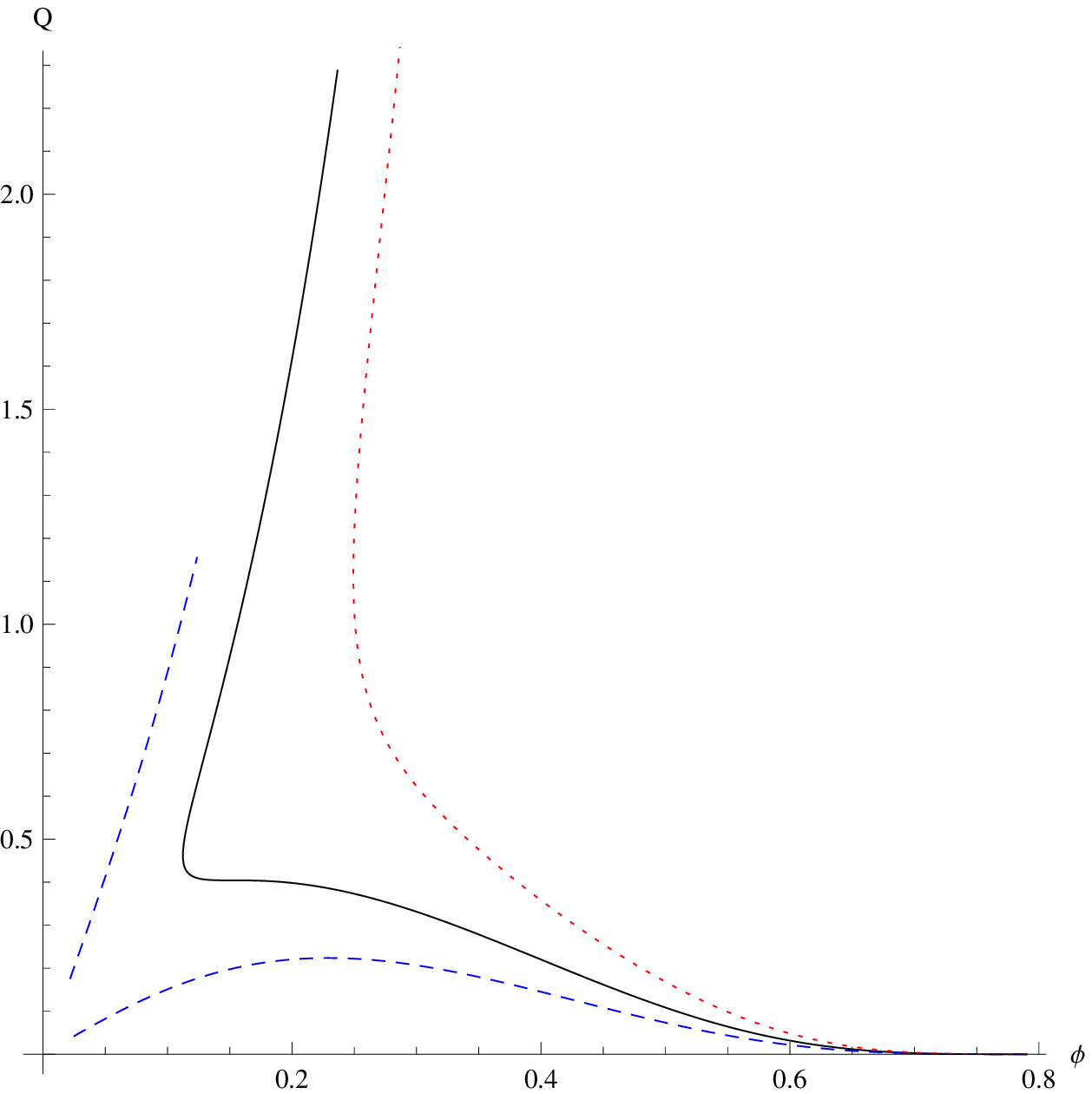}}
\caption[]{\it The $Q-\Phi$ curves for $d=4, 6, 7$ respectively. From top to the bottom the curves correspond to the effective temperature
$0.7T_{loc}$, $T_{loc}$, $1.4T_{loc}$.}}}
\label{FT3}
\end{figure}

When taking the $r_B$ and $T_{loc}$ as constant, we select the dual independent
variables $(\sigma, A)$, the critical position is given by the following conditions,
\ba
\label{eq18}
\left( {\frac{\partial A}{\partial \sigma }} \right)_{T_{loc} ,Q} &=&
\frac{1}{8\pi r_B }\frac{\left( {\frac{\partial T_{loc} }{\partial r_ + }}
\right)_{r_B } }{\left( {\frac{\partial \sigma }{\partial r_B }} \right)_{r_
+ } \left( {\frac{\partial T_{loc} }{\partial r_ + }} \right)_{r_B } -
\left( {\frac{\partial T_{loc} }{\partial r_B }} \right)_{r_ + } \left(
{\frac{\partial \sigma }{\partial r_ + }} \right)_{r_B } } = \frac{\left(
{\frac{\partial T_{loc} }{\partial r_ + }} \right)_{r_B } }{\frac{\partial
(\sigma ,T_{loc} )}{\partial (r_B ,r_ + )}} = f(r_ + ,r_B ) = 0, \no \\
\left( {\frac{\partial ^2A}{\partial \sigma ^2}} \right)_{T_{loc} ,Q} &=&
\left( {\frac{\partial f(r_ + ,r_B )}{\partial \sigma }} \right)_{T_{loc}
,Q} = 0.
\ea
Comparing Eqs.~\ref{eq14},~\ref{eq17} and~\ref{eq18}, we can find that
the equation of critical point is same. So we can obtain the same critical
point with different independent dual variable in the thermodynamics system.
The critical point is unaltered with the choose for independent dual variables.

According to Ehrenfest's classification, when the chemical potential and its first derivative are continuous, whereas
the second derivative of chemical potential is discontinuous, this kind of
phase transition is called the second-order phase transition.
For Van der Waals system there is no latent heat and the liquid-gas structure do not change suddenly at the critical
point. Therefore this kind of phase transition belongs to the second-order phase transition and continuous phase
transition. To discuss the critical behaviors of system near the
phase transition point with $r_B$ is unaltered, we will evaluate the second
order partial derivation of chemical potential,
\ba
\label{eq25}
\alpha &=& \frac{1}{\Phi }\left( {\frac{\partial \Phi }{\partial T_{loc} }}
\right)_Q = \frac{1}{\Phi }\left( {\frac{\partial \Phi }{\partial r_ + }}
\right)_Q \left( {\frac{\partial T_{loc} }{\partial r_ + }} \right)_Q^{ - 1}
, \no \\
k_T &=& - \frac{1}{\Phi }\left( {\frac{\partial \Phi }{\partial Q}}
\right)_{T_{loc} } = - \frac{1}{\Phi }\left[ {\left( {\frac{\partial \Phi
}{\partial Q}} \right)_{r_ + } \left( {\frac{\partial T_{loc} }{\partial r_
+ }} \right)_Q - \left( {\frac{\partial \Phi }{\partial r_ + }} \right)_Q
\left( {\frac{\partial T_{loc} }{\partial Q}} \right)_{r_ + } }
\right]\left( {\frac{\partial T_{loc} }{\partial r_ + }} \right)_Q^{ - 1} ,
\no \\
C_Q &=& T_{loc} \left( {\frac{\partial S}{\partial T_{loc} }} \right)_Q =
T_{loc} \left( {\frac{\partial S}{\partial r_ + }} \right)\left(
{\frac{\partial T_{loc} }{\partial r_ + }} \right)_Q^{ - 1} .
\ea
From Eqs.~\ref{eq14},~\ref{eq17} and~\ref{eq18}, we know that these quantities $\alpha$, $k_T$ and $C_Q$, which obtained in Eq~\ref{eq25},
approach to infinity at the critical point. So the phase transition, which happens in the black hole
thermodynamics ensemble, is an continuous phase transition. The Gibbs free energy and the first order
partial derivation for these two phase is continuous at the critical point of continuous phase.
\begin{figure}[!htbp]
\center{\subfigure[~$ \alpha-r_+$] {
\includegraphics[angle=0,width=5cm,keepaspectratio]{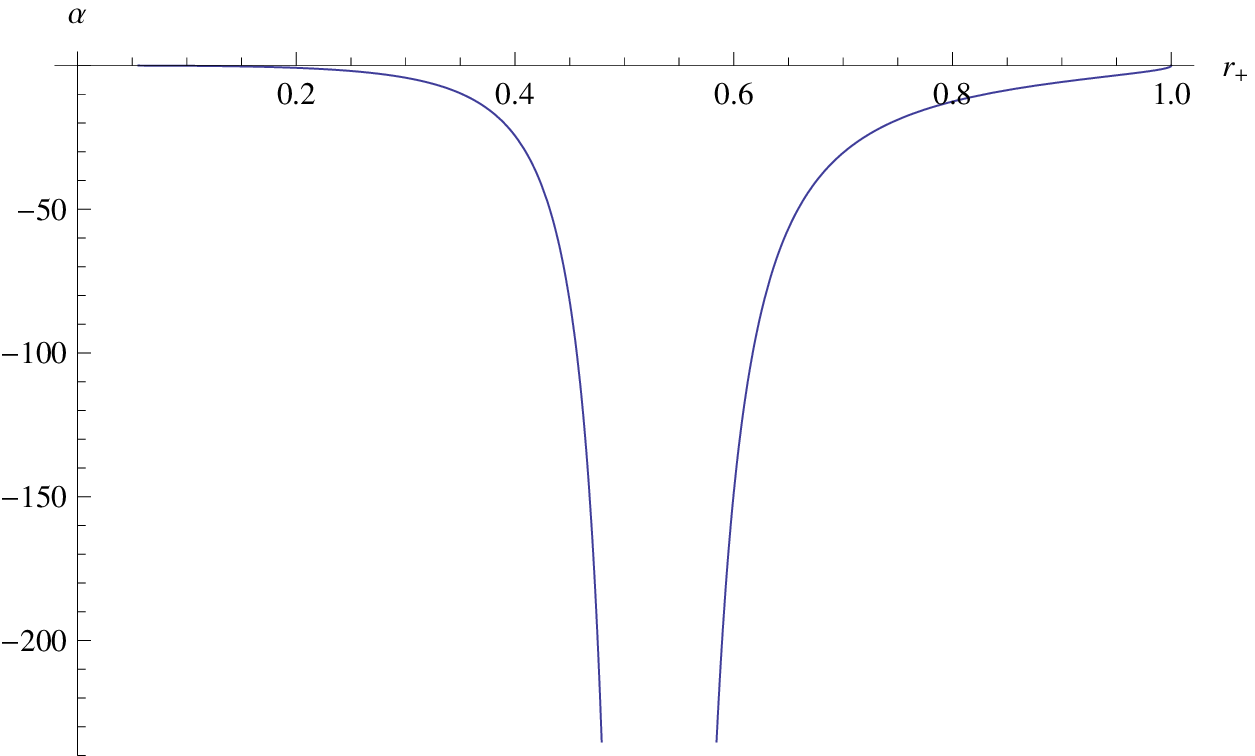}}
\subfigure[~$ \kappa-r_+$] {
\includegraphics[angle=0,width=5cm,keepaspectratio]{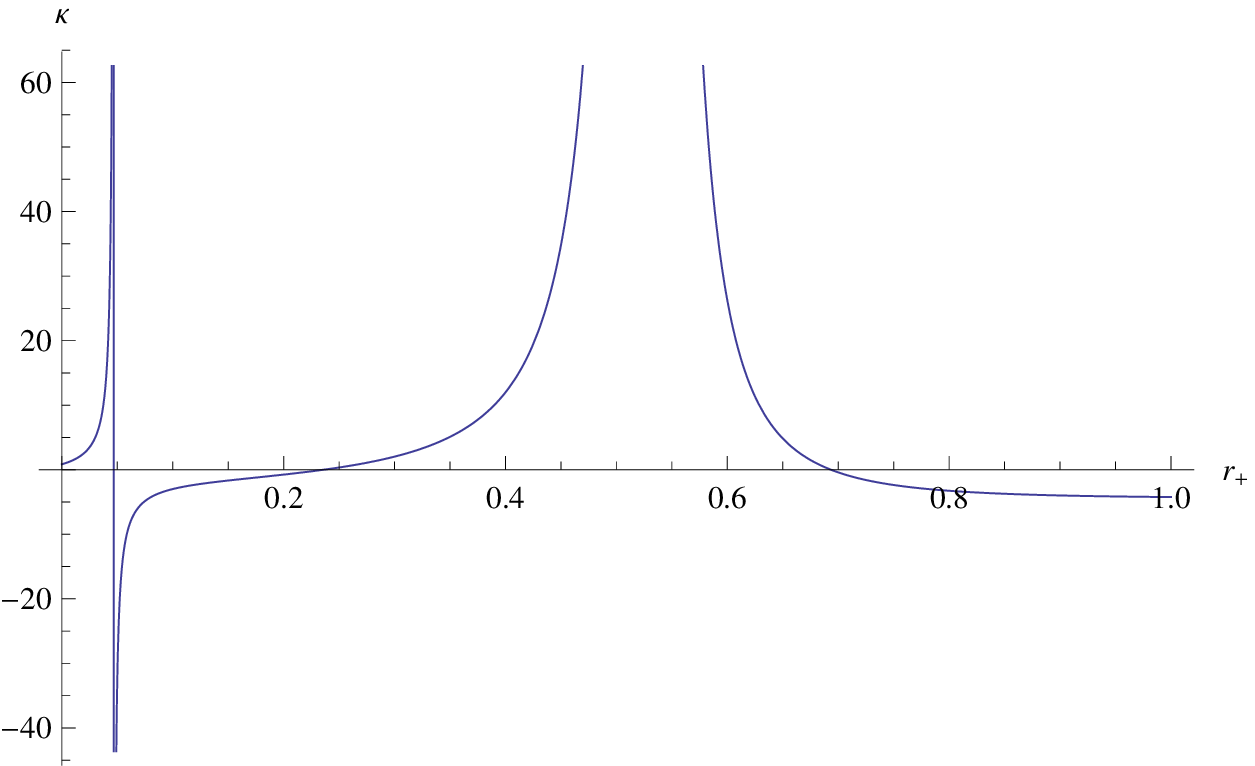}}
\subfigure[~$ C_Q-r_+$] {
\includegraphics[angle=0,width=5cm,keepaspectratio]{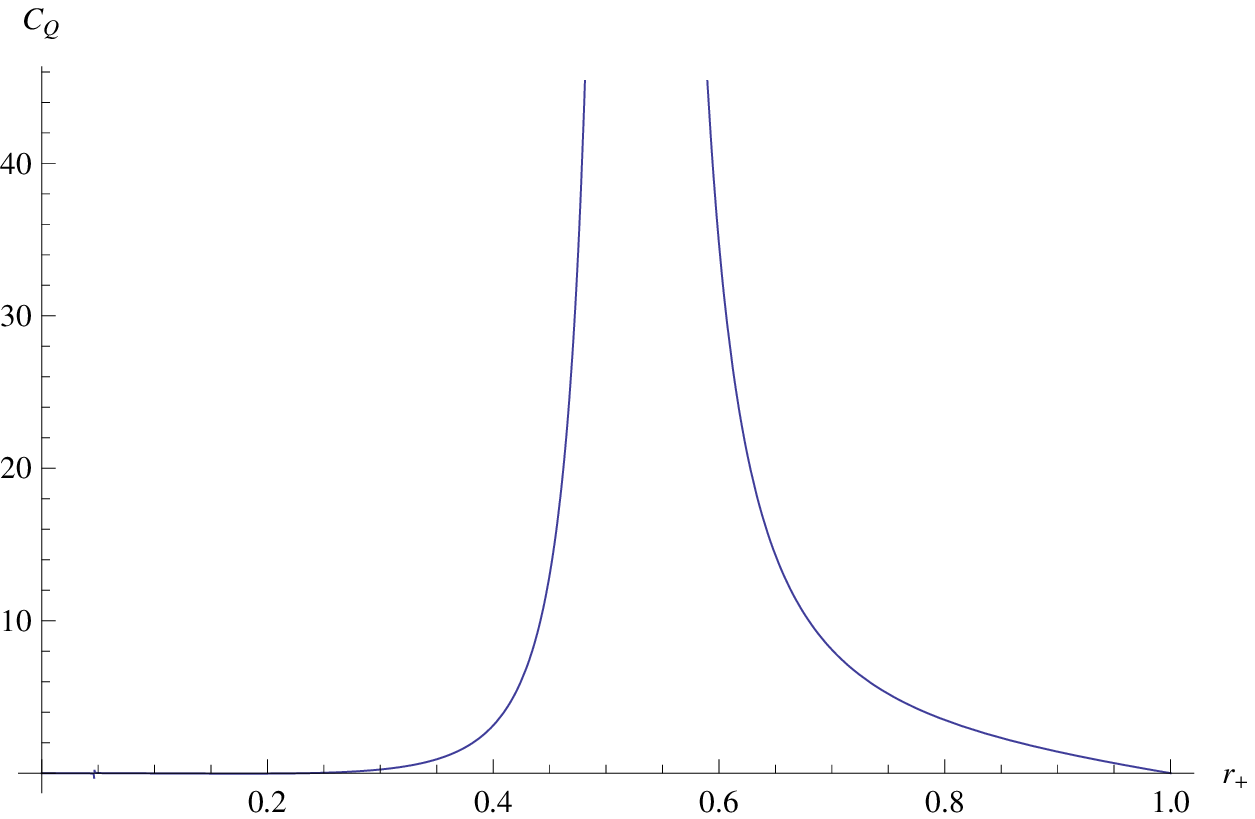}}
\caption[]{\it $\alpha-r_+$, $\kappa-r_+$ and $C_Q-r_+$ curves for higher dimensional charged black hole with $d=4$}}
\label{Cx}
\end{figure}
We also depict the curves of $\alpha-r_+$, $\kappa-r_+$ and $C_Q-r_+$ in the Fig. 5 respectively at the constant pressure.
From these curves, we find that the specific heat of black hole thermodyanmic system at constant charge $C_Q $,
the expansion coefficient $\alpha $, and the compressibility $\kappa $
have infinite peak.

For usual thermodynamic system, the entropy change and volume change are equivalent respectively, as $dv^1 = dv^2$
and $dS^1 = dS^2$, near critical point, Ehrenfest had calculated the Enrenfest equation,
\ba
\label{eq32}
\frac{dP}{dT} = \frac{\alpha ^1 - \alpha ^2}{k_T^1 - k_T^2 },
\quad
\frac{dP}{dT} = \frac{C_P^1 - C_P^2}{Tv(\alpha ^1 - \alpha ^2)}.
\ea
in which the subscript 1 and 2 represent phase 1 and 2 respectively.
Recently, the Ehrenfest equation for AdS black hole thermodynamics
system~\cite{JXM2,JXM3,JXM4,JXM5,JXM6,zhao7,Jing,Banerjee}, has been studied and the PD
Prigogine-Defay(PD) relation is obtained
\ba
\Pi = \frac{\Delta C_P \Delta k_T }{TV(\Delta \alpha )^2} = 1.
\ea
For the black hole thermodynamic system, we can rewrite the equation of Ehrenfest as
\ba
\begin{split}
\label{eq39}
\left( {\frac{\partial Q}{\partial T_{loc} }} \right)_\Phi &= \frac{\alpha _2
- \alpha _1 }{k_{T_2 } - k_{T_1 } } = \frac{\Delta \alpha }{\Delta k_T },
\\
\left( {\frac{\partial Q}{\partial T_{loc} }} \right)_S &= \frac{C_{Q_2 } -
C_{Q_1 } }{T_{loc} \Phi (\alpha _2 - \alpha _1 )} = \frac{\Delta C_Q
}{T_{loc} \Phi \Delta \alpha }.
\end{split}
\ea
From
\ba
\label{eq40}
\left( {\frac{\partial T_{loc} }{\partial Q}} \right)_\Phi = \left(
{\frac{\partial T_{loc} }{\partial Q}} \right)_S - \left( {\frac{\partial
T_{loc} }{\partial S}} \right)_Q \left( {\frac{\partial \Phi }{\partial Q}}
\right)_S \left( {\frac{\partial \Phi }{\partial S}} \right)_Q^{ - 1} .
\ea
According to Eq. (\ref{eq17}), when, the critical points
satisfy
\be
\label{eq43}
\left( {\frac{\partial T_{loc} }{\partial r_ + }} \right)_Q = \left(
{\frac{\partial T_{loc} }{\partial S}} \right)_Q \left( {\frac{\partial
S}{\partial r_ + }} \right)_Q = 0.
\ee
From Eq. (\ref{eq5}), we know
\begin{equation}
\label{eq44}
\left( {\frac{\partial S}{\partial r_ + }} \right)_Q = \frac{\omega _{d - 2}
(d - 2)r_ + ^{d - 3} }{4} \ne 0.
\end{equation}
So, Eq.(\ref{eq40}) can be written as
\begin{equation}
\label{eq45}
\left( {\frac{\partial T_{loc} }{\partial Q}} \right)_\Phi ^c = \left(
{\frac{\partial T_{loc} }{\partial Q}} \right)_S^c .
\end{equation}
Substituting Eq .(\ref{eq45}) to Eq .(\ref{eq39}), the Prigogine--Defay (PD) ratio ($\Pi
)$ can be calculated as
\begin{equation}
\label{eq46}
\Pi =\frac{\Delta C_Q \Delta \kappa _T }{T_{loc} \Phi(\Delta \alpha )^2}=1.
\end{equation}
Hence, when $Q$ is constant, the phase transition occurred at $T_{loc} =T_{loc}^c $ is a second
order equilibrium transition. This is true in spite of the
fact that the phase transition curves are smeared and divergent near the
critical point. This result is in agreement with the result of the AdS black holes.

\section{Discuss and Conclusions}

In this work, to study the properties of black hole in asymptotically flat
spacetime, we built a stable black hole thermodynamic ensemble in the view of
thermodynamics. we calculated the local energy and local temperature, and find that these state parameters satisfy
the first law of thermodynamic Eq.(\ref{eq11}). Based on this condition, we analyze the critical behavior
of black hole thermodynamic ensemble through taking the state parameters $(Q,\Phi)$ of black hole
thermodynamic ensemble corresponding to the state parameters $(P,V)$ of Van der Waals system respectively.
we obtain the critical point of black hole thermodynamic system, and find the critical point is
independent of the dual independent variables we selected. This result for asymptotically flat
space is consistent with the conclusion for AdS spacetime~\cite{RBM}, and is an intrinsic property of
black hole thermodynamic ensemble. Firstly, From Table 1 we can find that the critical temperature
and critical charge of black hole ensemble increase with the spacetime dimensional increase, which is similar
to the critical behavior of AdS black hole. Secondly, the second partial derivative $\alpha, \kappa_T, C_Q$ of Gibbs free energy
of black hole thermodynamic ensemble is divergent at the critical point, the result is the same to
the result of AdS spacetime. Finally, the Prigogine-Defay(PD) relation at critical point in black hole
thermodynamic ensemble is consistent with the AdS black hole. The PD ration satisfies Eq.(\ref{eq46})
and is independent with spacetime dimension.

In the black hole thermodynamic ensemble, we built a concentric spherical
cavity, whose radii is fixed and larger than the ones of black hole, to achieve the
condition of thermodynamic stability for black hole ensemble. The reason, which AdS black hole
can achieve the thermodynamic stability, is that there is a cosmology constant $\Lambda$. Through comparing the
critical behavior of thermodynamic ensemble with the ones of AdS spacetime, we can argue that
the radii $r_B$ of concentric spherical cavity is related to the cosmology constant $\Lambda$ in the thermodynamic
view.

\begin{acknowledgments}\vskip -4mm
We would like to thank Dr Meng-Sen Ma and
Hui-Hua Zhao for their indispensable discussions and comments.
H. F. Li would like to appreciate the Program for the Innovative Talents of Higher
Learning Institutions of Shanxi and are also grateful to KITPC for its hospitality and its partial support during the
completion of this work. This work was supported in part by the National Natural Science
Foundation of China (Grant Nos. 11075098, 11175109,11475108), by the Young
Scientists Fund of the National Natural Science Foundation of China
(Grant No.11205097), the Natural Science Foundation for Young
Scientists of Shanxi Province,China (Grant No.2012021003-4), the Natural Science
Foundation of Datong city(Grant No.20150110).
\end{acknowledgments}

\end{CJK*}

\end{document}